\documentclass[reprint,aps,prl,twocolumn,superscriptaddress,showpacs,floatfix,preprintnumbers]{revtex4-1}
\usepackage{graphicx}
\usepackage{color}
\usepackage{natbib,color,bm}
\usepackage{multirow}
\usepackage{amsmath}

\def\TN{\ensuremath{T_{\rm N}}}
\def\Hc{\ensuremath{H_{\rm c}}}
\def\LYF{LiYbF$_4$}
\def\LEF{LiErF$_4$}
\def\kk{\ensuremath{\mathbf k}}

\begin{document}

\title{Dimensional Reduction in Quantum Dipolar Antiferromagnets}

\author{P. Babkevich}
\email{peter.babkevich@epfl.ch}
\affiliation{Laboratory for Quantum Magnetism, Institute of Physics, \'{E}cole Polytechnique F\'{e}d\'{e}rale de Lausanne (EPFL), CH-1015 Lausanne, Switzerland}
\author{M. Jeong}
\affiliation{Laboratory for Quantum Magnetism, Institute of Physics, \'{E}cole Polytechnique F\'{e}d\'{e}rale de Lausanne (EPFL), CH-1015 Lausanne, Switzerland}
\author{Y. Matsumoto}
\affiliation{Institute for Solid State Physics, University of Tokyo, Kashiwa, Chiba 277-8581, Japan}
\author{I. Kovacevic}
\affiliation{Laboratory for Quantum Magnetism, Institute of Physics, \'{E}cole Polytechnique F\'{e}d\'{e}rale de Lausanne (EPFL), CH-1015 Lausanne, Switzerland}
\author{A. Finco}
\affiliation{Laboratory for Quantum Magnetism, Institute of Physics, \'{E}cole Polytechnique F\'{e}d\'{e}rale de Lausanne (EPFL), CH-1015 Lausanne, Switzerland}
\affiliation{ICFP, D\'{e}partement de physique, \'{E}cole normale sup\'{e}rieure, 45 rue d'Ulm,
75005 Paris, France}
\author{R.~Toft-Petersen}
\affiliation{Helmholtz-Zentrum Berlin f\"ur Materialien und Energie, D-14109 Berlin, Germany}
\author{C. Ritter}
\affiliation{Institut Laue-Langevin, BP 156, F-38042, Grenoble Cedex 9, France}
\author{M. M{\aa}nsson}
\affiliation{Laboratory for Quantum Magnetism, Institute of Physics, \'{E}cole Polytechnique F\'{e}d\'{e}rale de Lausanne (EPFL), CH-1015 Lausanne, Switzerland}
\affiliation{Laboratory for Neutron Scattering, Paul Scherrer Institut, CH-5232 Villigen, Switzerland}
\affiliation{Department of Materials and Nanophysics, KTH Royal Institute of Technology, SE-164 40 Kista, Sweden}
\author{S. Nakatsuji}
\affiliation{Institute for Solid State Physics, University of Tokyo, Kashiwa, Chiba 277-8581, Japan}
\author{H. M. R\o nnow}
\affiliation{Laboratory for Quantum Magnetism, Institute of Physics, \'{E}cole Polytechnique F\'{e}d\'{e}rale de Lausanne (EPFL), CH-1015 Lausanne, Switzerland}

\begin{abstract}
We report ac susceptibility, specific heat and neutron scattering measurements on a dipolar-coupled antiferromagnet \LYF. For the thermal transition, the order-parameter critical exponent is found to be 0.20(1) and the specific-heat critical exponent $-0.25(1)$. The exponents agree with the 2D XY$/h_4$ universality class despite the lack of apparent two-dimensionality in the structure. The order-parameter exponent for the quantum phase transitions is found to be 0.35(1) corresponding to $(2+1)$D. These results are in line with those found for \LEF\ which has the same crystal structure, but largely different \TN, crystal field environment and hyperfine interactions. Our results therefore experimentally establish that the dimensional reduction is universal to quantum dipolar antiferromagnets on a distorted diamond lattice.
\end{abstract}

\maketitle

Critical phenomena near continuous phase transitions do not depend on the microscopic details of systems but only on the symmetry of the order parameter and interactions and the spatial dimensionality \cite{stanley1987introduction}. Such universality for classical thermal transitions has been thoroughly demonstrated with various physical systems over decades while nowadays a similar line of effort is actively pursued for zero-temperature quantum transitions \cite{sondhi1997continuous, vojta2003quantum, sachdev2007quantum}. Comparing experimental observations with theoretical models has been particularly successful for magnetic insulators that could be simply modeled by short-ranged, exchange-coupled spins on a lattice. Although dipolar interactions appear to be more classical than their exchange-coupled counterparts, it has been shown that on a square or diamond lattice, quantum fluctuations can map long-ranged dipolar interactions to a two-dimensional Ising model \cite{shender-jetp-1982, henley-prl-1989, orendacova-cz-2002}. The Li$R$F$_4$ family is special as the rare-earth ions are arranged in a slightly distorted diamond-like structure making them intriguing to study in relation to order by disorder phenomena \cite{schustereit-crystals-2011}.

For the case of a dipolar-coupled Ising ferromagnet, the theoretical upper critical dimension D$^\ast=3$ and the mean-field calculations actually apply quite well as shown, for instance, in LiHoF$_4$ \cite{bitko-prl-1996}. This is despite the significant role of hyperfine interactions around the quantum phase transition \cite{ronnow-science-2005, kovacevic-submitted}. Recently, quantum and classical critical properties of a long-range, dipolar-coupled antiferromagnet could be investigated for the first time with \LEF\ \cite{kraemer-science-2012}. It was discovered that the specific-heat and order-parameter critical exponents, $\alpha=-0.28(4)$ and $\beta_T=0.15(2)$, for the thermal transition are totally different from the mean-field predictions of $\alpha=0$ and $\beta_T=0.5$. Instead, these exponent values suggest a 2D XY$/h_4$ universality class, despite the absence of any apparent two-dimensionality in the structure of the system. This intriguing dimensional reduction was further corroborated by the $\beta_H=0.31(2)$ for the quantum transition induced by applying a longitudinal magnetic field, which corresponds to $(2+1)$D, as expected from quantum-classical mapping \cite{sachdev2007quantum}. Whether the dimensional reduction is universal to all dipolar quantum antiferromagnets or is special to \LEF, due to rather close (3\,meV) higher-lying crystal-field levels or weak hyperfine interactions, is to date unknown.

Among the Li$R$F$_4$ family where $R$ is a rare-earth ion, \LYF\ has been suggested to be an alternate candidate for a dipolar antiferromagnet \cite{babkevich-prb-2015}. However, there are marked differences between \LYF\ and \LEF. First, the electronic level scheme is quite different with crystalline electric field split first excited state an order of magnitude higher in \LYF. Second, in Yb$^{3+}$, there are two stable isotopes of Yb with strong hyperfine coupling -- $11.0\,\mu$eV for $^{171}$Yb (14.3\%) and  $-3.0\,\mu$eV for $^{173}$Yb (16.1\%). \LEF\ contains $^{167}$Er (22.8\%) whose hyperfine coupling strength is weak, 0.5\,$\mu$eV. Therefore, \LYF\ could serve as an excellent candidate to test for the robustness of dimensional reduction in dipolar antiferromagnets arranged on a distorted diamond lattice.

\begin{figure}[t]
\includegraphics[width=1\columnwidth]{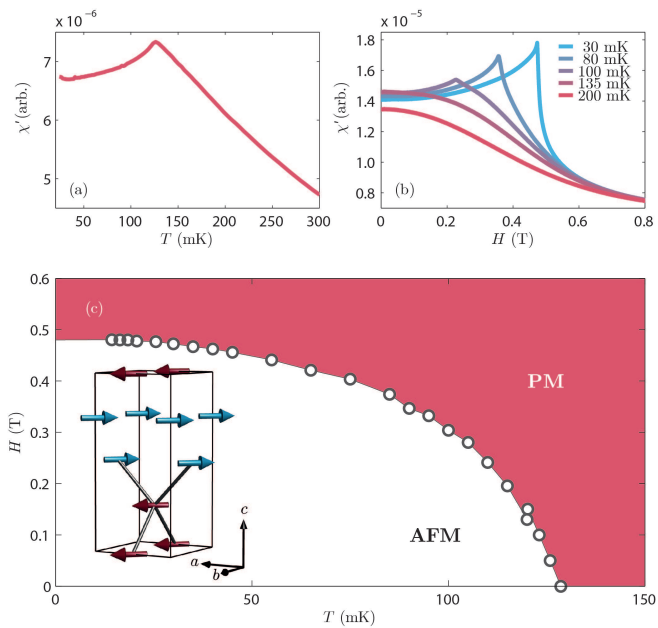}
\caption{(a) Real part of ac susceptibility $\chi'$ as a function of temperature in zero field and (b) $\chi'$ as a function of field at different temperatures. (c) Magnetic phase diagram mapped out using the susceptibility. Inset shows the bilayer magnetic structure of \LYF.}
\label{fig1}
\end{figure}

In this Letter, we present ac susceptibility, specific heat, and neutron scattering measurements on \LYF\ and demonstrate the thermal and quantum critical properties. The field-temperature ($H$-$T$) phase diagram is first mapped out and a bilayered XY antiferromagnetic order for the ground state is identified. Then we show that the critical exponents $\alpha$, $\beta_T$, and $\beta_H$ support the dimensional reduction as a universal feature of quantum dipolar antiferromagnets.

Large, high-quality single crystals were obtained from a commercial source. In order to reduce neutron absorption, the samples were enriched with the $^7$Li isotope. The ac susceptibility $\chi(T,H)$ was measured on a single crystal using mutual inductance method where the excitation field was 40\,mOe and the excitation frequency 545\,Hz. The specific heat $C_p(T)$ was measured by the relaxation method in a dilution refrigerator with a temperature stability of 0.1\,mK. Powder neutron diffraction was performed using the high-intensity D1B and high-resolution D2B diffractometers at ILL, France using incident neutron wavelength 2.52 and 1.59\,\AA, respectively. The evolution of the magnetic Bragg peak intensities with temperature and field was followed by performing high-resolution single-crystal neutron scattering using the triple-axis spectrometer FLEXX at HZB, Germany \cite{le-flexx}. The instrument was set up with 40' collimation before and after the sample and incident neutron wavelength of $\lambda = 4.05$\,\AA. The corresponding wavevector and energy resolution (FWHM) was on the order of 0.014\,\AA\ and 0.15\,meV, respectively.

Figure~\ref{fig1} shows bulk ac susceptibility data from a single-crystal \LYF. The temperature-field phase boundary was mapped for a transverse magnetic field applied along the $c$ axis. Figure~\ref{fig1}(a) shows the real part of the ac susceptibility, $\chi'$, as a function of temperature in zero field. The peak in zero field reflects the antiferromagnetic transition at $\TN=130$\,mK. Figure~\ref{fig1}(b) shows $\chi'(H)$ at 30-200\,mK. Below \TN, a pronounced cusp is observed which corresponds to a quantum transition from the ordered to a quantum paramagnetic phase. At base temperature, a maximum in $\chi'(H)$ is found at $\Hc=0.48$\,T. The peak shifts to lower fields as temperature is increased. Based on these measurements, we can accurately map out the phase diagram shown in Fig.~\ref{fig1}(c).

\begin{figure}[b]
\includegraphics[width=0.9\columnwidth]{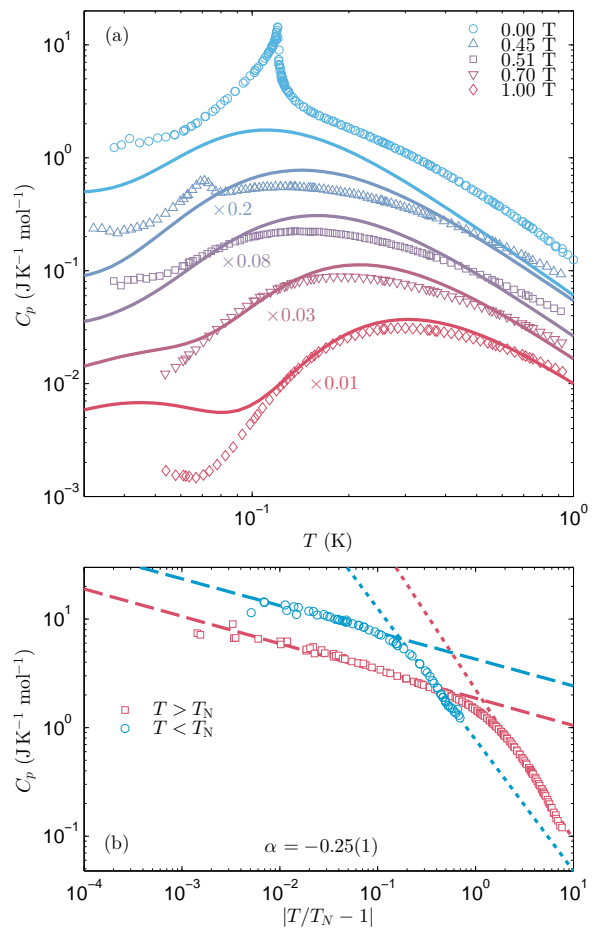}
\caption{(a) Specific heat in zero and finite fields as a function of temperature. Calculation of specific heat capacity in the single-ion limit for different fields are plotted by continuous lines. The data were displaced vertically by multiplying with scaling factors given in the figure. (b) Determination of the specific-heat critical exponent $\alpha$ for the thermal transition based on measurements above and below \TN\ (dashed line). Scaling away from the critical region was fitted by the dotted line.}
\label{fig2}
\end{figure}


The specific heat as a function of temperature is shown in Fig.~\ref{fig2}(a). In zero field, a sharp peak in the specific heat capacity marks the second-order thermal transition
\footnote{We note that the different techniques gave slightly different values of \TN, within 10\,mK. We attribute this to differing thermometer calibrations and possibly small thermal gradients between sample and thermometer. This does not affect the extracted exponents nor the main conclusions of this Letter.
}.
On applying a transverse field, we find the peak at \TN\ decreases in amplitude and shifts to lower temperature at $H=0.45$\,T. Above \Hc, only a broad hump is found in the specific heat capacity. At such low temperatures, phonon and crystal-field-level contributions are frozen out. We model the specific heat capacity away from the QPT using a parameter-free model where the Hamiltonian $\mathcal{H}$ contains crystal field, hyperfine and Zeeman terms. From the diagonalized Hamiltonian $\langle n | \mathcal{H} | n \rangle = \epsilon_n$, we calculate for each isotope $i$ the Schottky specific heat,
$C^{\rm Sch}_i = k_{\rm B}\beta^2\left[\langle\epsilon^2\rangle
 - \langle\epsilon\rangle^2\right]$,
where $k_{\rm B}$ is the Boltzmann factor and $\beta = 1/(k_{\rm B} T)$. The thermal ensemble average is denoted by $\langle \ldots \rangle$. The total specific heat capacity is found from the weighted sum of contributions from each Yb isotope.
%
%
The comparison between the experiment and our simple model is remarkably good considering that this is a parameter-free calculation with all parameters fixed from other experiments. It is possible to improve the comparison by including quadrupolar operators, and by fine-tuning hyperfine coupling strengths and the crystal field parameters, etc. However, this would give too many adjustable parameters, and the calculation anyway ignores collective effects beyond the mean-field level.

In zero applied field, close to \TN, the heat capacity can be described by a universal power-law,
\begin{equation}
C^{\rm crit}_p = A|t|^{\alpha} + B,
\end{equation}
where the reduced temperature $t = 1-T/\TN$, $A$ and $B$ are free parameters which can have different values above and below \TN. The results of our analysis are shown in Fig.~\ref{fig2}(b). The contribution from the background term, $B$, is found to be small and is set to zero above and below \TN. A good fit is found for $\alpha=-0.25(1)$, similar to the value of $-0.28(4)$ found in \LEF\ \cite{kraemer-science-2012}. The negative exponents imply that $C_p$ is finite at \TN.
Away from the phase transition we observe a change in the scaling. Above around 250\,mK and below 100\,mK the data can be fit to an exponent of around $-1.3(1)$. It is somewhat surprising that the critical scaling can be traced out all the way to 2\TN\ and is dramatically different to \LEF\ where a cross over was found above 1.03\TN\ \cite{kraemer-science-2012}.

\begin{figure}
\includegraphics[width=\columnwidth]{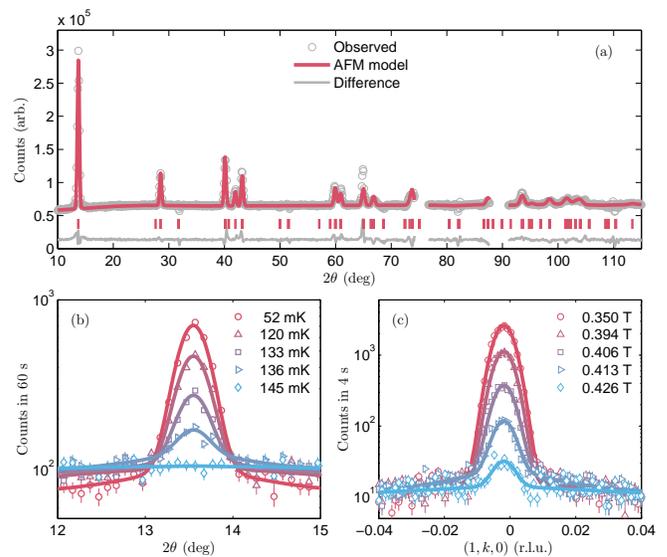}
\caption{(a) Magnetic powder diffraction pattern from the subtraction of paramagnetic background from 50\,mK measurements. (b) Magnetic Bragg peak from powder diffraction at different temperatures in zero field and (c) single-crystal measurements at 70 mK in different fields. Lines are fits to a Gaussian with additional contribution from critical scattering.}
\label{fig3}
\end{figure}


To elucidate the magnetic structure below \TN, we performed neutron diffraction on a powder of \LYF. At 10\,K, in the paramagnetic phase, the crystal lattice was refined using the $I4_1/a$ space group where $a=5.13433(8)$\,\AA\ and $c=10.5917(2)$\,\AA. Below 140\,mK we find additional peaks which emerge from antiferromagnetic ordering corresponding to a $\mathbf{k}=(1,0,0)$ magnetic propagation wavevector. Figure~\ref{fig3}(a) shows powder diffraction pattern obtained by subtracting measurements above \TN\ from 50\,mK data. The magnetic peaks are well described by a bilayer antiferromagnetic structure with moments along the [110] direction, where moments related by $I$-centering are aligned antiparallel. An ordered moment of 1.9(1)$\mu_{\rm B}$ is found to reside on each Yb$^{3+}$ ion. A schematic of a possible magnetic structure is shown in Fig.~\ref{fig1}(c). This differs from \LEF\ where the moments are parallel to the [100] direction. Although our data do not allow us to uniquely identify the magnetic structure, it is clear that \LEF\ and \LYF\ order differently (see Supplemental Material). The origin of this is not entirely obvious but could be attributed to the in-plane anisotropy set by the crystal field. This would depend primarily on the $B_4^4(c)\mathbf{O}_4^4(c)$ crystal field term and result in the configuration energy $E\sim B_4^4(c) \cos(4\phi)$ having minima rotated by 45$^\circ$ when changing the sign of $B_4^4(c)$ parameter. Indeed, our previously reported results show that $B_4^4(c)$ is significantly larger and of opposite sign in \LYF\ compared to \LEF\ \cite{babkevich-prb-2015}.

The powder sample of \LYF\ was measured as a function of temperature in fine steps across the thermal phase transition. Figure~\ref{fig3}(b) shows how the magnetic intensity of the $(001)$ reflection decreases with temperature. As expected from ac susceptibility and heat capacity measurements, magnetic order disappears above 136\,mK.
Single-crystal measurements as a function of transverse field are shown in Fig.~\ref{fig3}(c). At $T_{\rm base} = 70$\,mK, a field of around 0.43\,T suppresses the $(100)$ magnetic peak. A small contribution from critical scattering is observed as tails of the main peak. The neutron scattering measurements of \LYF\ reaffirm the phase diagram found from ac susceptibility in Fig.~\ref{fig1}.

\begin{figure*}
\includegraphics[width=0.9\textwidth]{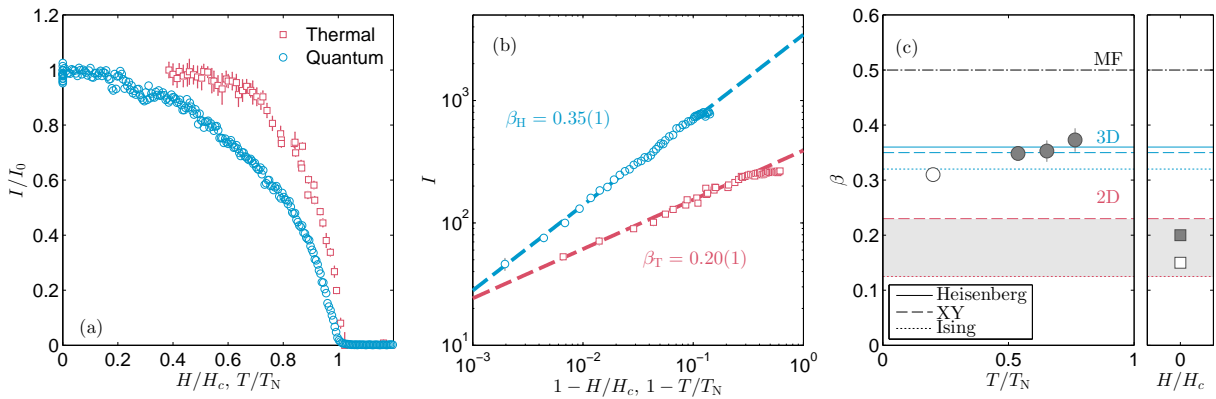}
\caption{(a) Evolution of the zero-field Bragg peak intensity as a function of temperature and the Bragg peak intensity as a function of the field at 50\,mK. (b) Determination of the order-parameter critical exponent for the thermal classical phase transition and the quantum transition.
(c) Extraction of the critical exponent $\beta_H$ at different temperatures is plotted by circles. The critical exponent $\beta_T$ is shown by squares in the panel on the right. Filled data points represent exponents found in this work for \LYF, in addition empty symbols denote \LEF\ results, after Ref.~\cite{kraemer-science-2012}.
Dashed lines and dotted horizontal lines correspond to critical exponents of 3D \cite{lovesey-book} and 2D \cite{kagawa-nature-2005, taroni-jpcm-2008} universality classes, respectively. The expected mean-field (MF) result of $\beta=0.5$ is also plotted.}
\label{fig4}
\end{figure*}

The evolution of the magnetic Bragg peak intensities with temperature and field are shown in Fig.~\ref{fig4}(a). Continuous onset and smooth evolution of the order parameter is observed with both temperature and field.

For both the powder and single-crystal data we have considered a model consisting of (i) a Lorentzian lineshape to describe the critical fluctuations close to the phase transition and (ii) a delta-function to account for long-range order. Both of these were then convoluted by a Gaussian, representing the instrumental resolution. The strength of scattering from critical fluctuations is rather weak and within the measured resolution and statistics cannot be refined to extract further exponents in either powder or single-crystal data. The amplitude of the convoluted delta-function $\sigma$ corresponds to the square of the order parameter, i.e., staggered magnetization. Therefore, sufficiently close to the phase boundary, $\sigma \propto t^{2\beta_T}$ for a zero-field measurement and $\sigma \propto h^{2\beta_H}$, where $h = 1-H/\Hc$ on sweeping magnetic field at constant temperature. From such treatment we obtain the results shown in Fig.~\ref{fig4}(b), where squares and circles are for thermal and quantum critical exponents, respectively. Fitting the data to a power-law, we obtain $\beta_T = 0.20(1)$ and $\beta_H = 0.35(1)$.

The base temperature of 70\,mK at which the field was swept to cross the quantum phase transition may appear rather high as $T_\mathrm{base}\simeq 0.5\, \TN$. For \LEF, on the other hand, the $\beta_H$ was extracted at $T_\mathrm{base} \simeq 0.2\, \TN$. To ensure that the extracted $\beta_H = 0.35(1)$ is correct and not affected by thermal fluctuations, we followed the field evolution of the $(100)$ Bragg peak at a few higher temperatures. We find, as shown in Fig.~\ref{fig4}(c), no appreciable change in $\beta_H$ in the temperature range studied. This assertion is further corroborated by the heat capacity measurements, shown in Fig.~\ref{fig2}, where the thermal critical region is found above around 0.8\TN. Comparing the critical exponents to tabulated results \cite{lovesey-book, kagawa-nature-2005, taroni-jpcm-2008}, it is clear that the quantum transition falls in the $\beta=0.32$-0.36 range predicted for 3D models. While the 2D XY/$h_4$ model predicts $\beta=0.125$-0.23, bound by 2D Ising and XY transitions, which best describes the thermal phase transition \cite{taroni-jpcm-2008}.

Such dimensional reduction has been hinted at from studies of other dipolar systems. A good example is $R$Ba$_2$Cu$_3$O$_{7-\delta}$ whose dipolar interactions were the focus of some theoretical work \cite{debell-jpcm-1991, macisaac-prb-1992}. It was argued two-dimensional behavior is strongly related to the spacing of basal planes with a cross-over from three-dimensional behavior around $c/a > 2.5$. However, relatively strong exchange coupling as well as superconductivity makes this system more complicated to separate the influence of the dipolar interaction. We hypothesize that systems such as $R$PO$_4$(MoO$_3$)$_{12}\cdot$30H$_2$O where rare-earth ions form a diamond lattice would also be a good candidate to examine quantum criticality due to strong dipolar and weak exchange interactions \cite{noordaa-jlow-2000}. Quantum spin fluctuations of dipolar-coupled antiferromagnetism have already been suggested to play a major role in these systems \cite{white-prl-1993}.


To conclude, dipolar-coupled \LYF\ undergoes a thermal transition into the bilayer, XY antiferromagnetically ordered phase, where the critical exponents follow the 2D XY$/h_4$ universality class despite the lack of apparent two-dimensionality in the structure. Applying a transverse magnetic field suppresses the order, inducing a quantum phase transition into a paramagnetic state, which scales according to $(2+1)$D universality. These observations are in accordance with those for \LEF\ with largely different crystal field environment, \TN, and hyperfine interactions. Our results, therefore, experimentally establish that the dimensional reduction is a universal feature of dipolar-coupled quantum antiferromagnets on the distorted diamond-like lattice and are likely to be applicable to a vast range of seemingly different systems. While it may be premature to conclude that dimensional reduction is universal to other lattices, the challenge is now to find a dipolar-coupled antiferromagnet without it.

\begin{acknowledgments}
We are grateful to B. Klemke for his technical support. We would like to thank J. O. Piatek for his help in setting up the ac susceptibility measurements and I. \v{Z}ivkovi\'{c} for helpful discussions. We are indebted to J.~S. White, M. Zolliker and M. Bartkowiak for their assistance during preliminary measurements on the TASP spectrometer at SINQ, PSI.
This work was funded by the Swiss National Science Foundation and its Sinergia network MPBH, Marie Curie Action COFUND (EPFL Fellows), and European Research Council Grant CONQUEST. This work is partially supported by Grants-in-Aid for Scientific Research (No. 25707030 and No. 15K13515) and Program for Advancing Strategic International Networks to Accelerate the Circulation of Talented Researchers (No. R2604) from the Japanese Society for the Promotion of Science. We thank HZB for the allocation of neutron radiation beam time. This project has received funding from the European Union's Seventh Framework Programme for research, technological development and demonstration under the NMI3-II Grant No. 283883. M. M. was partly supported by Marie Sklodowska Curie Action, International Career Grant through the European
Commission and Swedish Research Council (VR), Grant No. INCA-2014-6426.
\end{acknowledgments}

\bibliography{shorttitles,biblio_v0p5}

\newpage
\section{Supplemental Material}
\subsection{Crystallographic structure}

\begin{figure}[b]
\includegraphics[width=0.95\columnwidth]{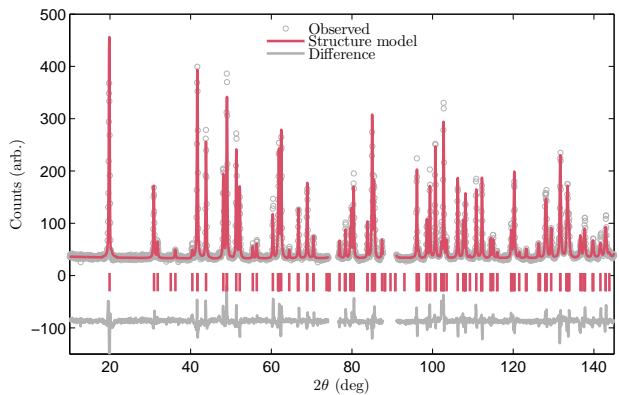}
\caption{(Color online) High-resolution neutron powder diffraction measurements using D2B diffractometer. Data collected at 10\,K and refined to the structural model described in the text. Incident neutron wavelength was 1.594\,\AA.}
\label{figS1}
\end{figure}

It is well known that systems of the Li$R$F$_4$ family crystallize in a scheelite CaWO$_4$ type structure. To verify our \LYF\ sample, we have performed careful measurements using D2B diffractometer in the paramagnetic phase at 10\,K -- well above magnetic ordering temperature. Our results are presented in Fig.~\ref{figS1}. A good fit to the diffraction pattern was found using Rietveld method in the Fullprof package \cite{carvajal-physicab-1993} which allows us to extract the atomic positions and $B_{\rm iso}$ isotropic Debye-Waller factors. In the case of $^7$Li, it was not possible to accurately refine the $B_{\rm iso}$ parameter and therefore it was fixed in the fitting. The detailed refinement, described in Table~\ref{tab:nuc_struct}, is in excellent agreement with that reported previously on the system in Ref.~\cite{thoma-inchem-1970}.



\begin{table}[b]
\centering
\begin{tabular}{l l l l l l}
\hline
\hline
Atom & site     & $x$ & $y$ & $z$ & $B_{\rm iso}$ (\AA$^2$)\\
\hline
$^7$Li & 4a & 0.0000       & 0.2500   & 0.1250  & 0.80  \\
Yb & 4b & 0.0000       & 0.2500   & 0.6250  & 0.09(3)  \\
F & 16f & 0.2186(4)       & 0.4169(4)   & 0.4571(2)  & 0.43(4)  \\
\hline
\hline
\end{tabular}
\caption{Nuclear structure refinement of \LYF\ shown in Fig.~\ref{figS1}. The Bragg peaks were indexed by $I4_1/a$ space group with lattice parameters of $a=5.13435(8)$\,\AA\ and $c=10.5918(2)$\,\AA. The fractional atomic positions using the second origin choice setting are listed in the table together with uncertainties given in brackets.
\label{tab:nuc_struct}}
\end{table}

\subsection{Magnetic structure}

\begin{figure*}[p]
\includegraphics[width=0.7\textwidth]{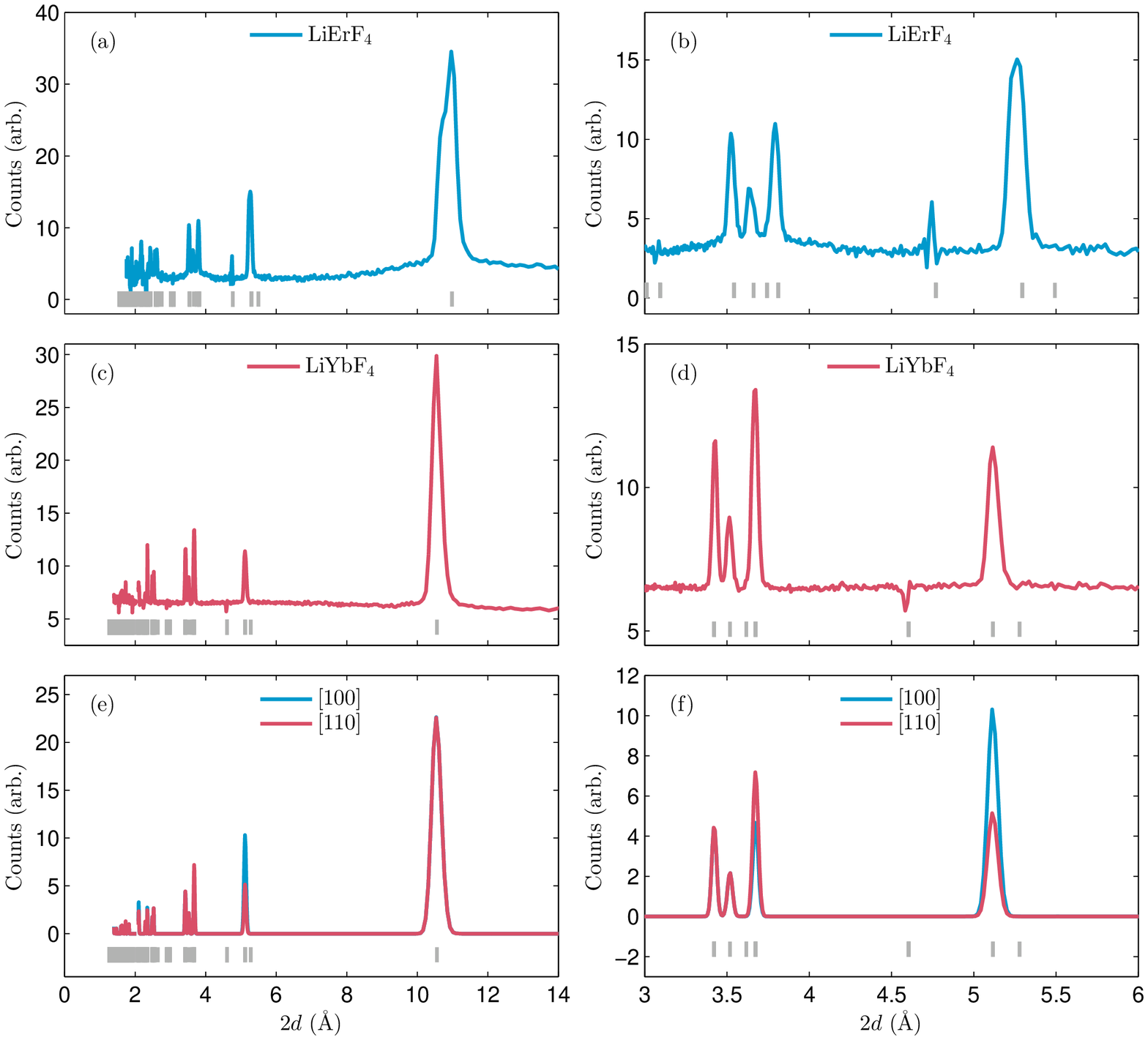}
\caption{(Color online) Neutron powder diffraction data recorded for (a,b) \LEF\ and (c,d) \LYF\ plotted as a function of $d$-spacing. In each case measurements in the paramagnetic phase were used to subtract the nuclear contribution to the patterns leaving purely the magnetic Bragg peaks. Grey vertical lines under the patterns show the indexation of the reflection. Simulations assuming collinear magnetic structures with moments along $[100]$ and $[110]$ directions are plotted in panels (e,f).}
\label{figS2}
\end{figure*}

Having confirmed the crystallographic structure of \LYF\ and the absence of impurities, we next consider the arrangement of the magnetic moments below \TN. Previous study of \LEF\ found that magnetic moments are arranged into a bilayer structure where the moments connected by $I$-centering are antiparallel \cite{kraemer-science-2012}. Indeed, solving the Hamiltonian in the mean-field approximation quickly converges to this structure. Our previous mean-field simulations of \LYF\ and \LEF\ indicate that the groundstate magnetic structures should be the same, with the only difference that the moment on Yb$^{3+}$ ion is expected to be smaller than that on Er$^{3+}$ \cite{babkevich-prb-2015}.

Neutron diffraction data from studies of \LEF\ is plotted in Figs.~\ref{figS2}(a) and (b). Measurements were collected using DMC diffractometer with $\lambda = 2.457$\,\AA. Antiferromagnetic ordering in \LEF\ sets in below 375\,mK \cite{kraemer-science-2012}. In order to obtain purely the magnetic contribution to the signal, we have subtracted measurements collected above 900\,mK. Surprisingly, some of the stronger peaks are found to sit on broad humps which could indicate some short-range correlations in the system but could also be some artifacts related to the background. The origin of these cannot be elucidated further.

In comparison, data collected using D1B at $\lambda = 2.52$\,\AA\ examining \LYF\ show a slowly varying background with no signs of any additional features. We notice from the \LEF\ and \LYF\ diffraction patterns shown in Figs.~\ref{figS2}(b) and (d) that the relative intensities of $(100)$ and $(102)$, close to 5.1 and 3.7\,\AA, respectively, are clearly different for the two systems. The ratio of $\sigma(100)$ to $\sigma(102)$ intensity in \LEF\ is 3.36(7) and in \LYF\ is 1.241(4).

Since the incident neutron wavelengths are very similar and the instrumental resolution is not very different for the two diffractometers we would have expected from mean-field simulations that the magnetic powder patterns are nearly the same. Intriguingly this does not appear to be the case. Performing Rietveld refinement of the magnetic structure for \LYF\ gives a better fit when the moments are allowed to rotate to be along the $[110]$ direction. The simulations for the two different moment directions is shown in Figs.~\ref{figS2}(e) and (f). In the model where the moments are along $[100]$, the $\sigma(100)/\sigma(102) = 4.14$  -- close to what we find for \LEF. Repeating this analysis for moments along $[110]$, we find instead $\sigma(100)/\sigma(102) = 1.35$, viz \LYF.

\subsection{Magnetic representation analysis}

\begin{table*}
\centering
\begin{tabular}{ l  c c c c  c c c c}
\hline
$\nu$
& $g_1$
& $g_2$
& $g_3$
& $g_4$
& $g_5$
& $g_6$
& $g_7$
& $g_8$ \\
\hline
1
& $\begin{pmatrix} 1 & 0  \\ 0 & 1 \end{pmatrix}$
& $\begin{pmatrix} 1 & 0  \\ 0 & 1 \end{pmatrix}$
& $\begin{pmatrix} 1 & 0  \\ 0 & -1 \end{pmatrix}$
& $\begin{pmatrix} 1 & 0  \\ 0 & -1 \end{pmatrix}$
& $\begin{pmatrix} 0 & 1  \\ 1 & 0 \end{pmatrix}$
& $\begin{pmatrix} 0 & 1  \\ 1 & 0 \end{pmatrix}$
& $\begin{pmatrix} 0 & -1  \\ 1 & 0 \end{pmatrix}$
& $\begin{pmatrix} 0 &-1  \\ 1 & 0 \end{pmatrix}$ \\
2
& $\begin{pmatrix} 1 & 0  \\ 0 & 1 \end{pmatrix}$
& $\begin{pmatrix} -1 & 0  \\ 0 & -1 \end{pmatrix}$
& $\begin{pmatrix} {\rm i} & 0  \\ 0 & -{\rm i} \end{pmatrix}$
& $\begin{pmatrix} -{\rm i} & 0  \\ 0 & {\rm i} \end{pmatrix}$
& $\begin{pmatrix} 0 & 1  \\ 1 & 0 \end{pmatrix}$
& $\begin{pmatrix} 0 & -1  \\ -1 & 0 \end{pmatrix}$
& $\begin{pmatrix} 0 & -{\rm i}  \\ {\rm i} & 0 \end{pmatrix}$
& $\begin{pmatrix} 0 & {\rm i}  \\ -{\rm i} & 0 \end{pmatrix}$ \\
\hline
\end{tabular}
\caption{Character table of the little group $G_\kk$ showing how the irreducible representations $\Gamma_\nu$ transform according to symmetry operations $g_1,\ldots,g_8$. Using the Seitz notation, the symmetry operations are defined as, $g_1 = \{1\mid 0,0,0 \}$, $g_2 = \{2_{00z} \mid 1/2,0,1/2\}$, $g_3 = \{4^+_{00z} \mid 3/4,1/4,1/4\}$, $g_4 = \{4^-_{00z} \mid 3/4,3/4,3/4\}$, $g_5 = \{-1\mid 0,0,0 \}$, $g_6 = \{m_{xy0} \mid 1/2,0,1/2\}$, $g_7 = \{-4^+_{00z} \mid 1/4,3/4,3/4\}$ and $g_8 = \{4^-_{00z} \mid 1/4,1/4,1/4\}$.
\label{tab:sym_elements}}
\end{table*}

\begin{table}
\centering
\begin{tabular}{ c  c  r r}
\hline
$\nu$ & $n$ & \quad$(\psi_x^1,\psi_y^1,\psi_z^1)$   & \quad$(\psi_x^2,\psi_y^2,\psi_z^2)$ \\
\hline
1 & 1 & $(1,0,0)$ & $(0,1,0)$ \\
1 & 2 & $(0,1,0)$ & $(-1,0,0)$ \\
1 & 3 & $(0,-1,0)$ & $(-1,0,0)$ \\
1 & 4 & $(1,0,0)$ & $(0,-1,0)$ \\
\hline
2 & 1 & $(1,0,0)$ & $(0,0,-{\rm i})$ \\
2 & 2 & $(0,0,{\rm i})$ & $(0,0,-1)$ \\
\hline
\end{tabular}
\caption{Basis functions $\psi$ of irreducible representation $\Gamma_\nu$ for ions situated at 1.~ $(x,y,z)$ and 2.~$(-y+3/4,x+1/4,z+1/4)$.
\label{tab:basis_vectors}}
\end{table}

The magnetic structures of \LYF\ and \LEF\ can be described by the magnetic propagation wavevector $\kk = (1,0,0)$. From the paramagnetic space group $I4_1/a$, the little group $G_\kk$ contains 8 symmetry elements ($g_1$ -- $g_8$) listed in Table~\ref{tab:sym_elements}. The magnetic representation $\Gamma_{\rm mag}$ of $G_\kk$ reduces to $\Gamma_{\rm mag} = 2\Gamma_1 + \Gamma_2$. Both $\Gamma_1$ and $\Gamma_2$ are two dimensional and their characters are given in Table~\ref{tab:sym_elements}. Using Basireps \cite{basireps}, we obtain basis functions $\psi$, shown in Table~\ref{tab:basis_vectors} for two symmetry-related sites. The two sites create an extinction condition which makes is possible to distinguish between magnetic moment directions even in the tetragonal cell with powder averaging. In general, the $n$th moment $\mathbf{m}_n$ can be expressed as a Fourier series,
\begin{equation}
\mathbf{m}_n = \sum_\kk \mathbf{S}_n^\kk e^{-{\rm i}\kk\cdot\mathbf{t}},
\end{equation}
where $\mathbf{t}$ is the real space translation vector. The vectors $\mathbf{S}_n^\kk$ are a linear sum of the basis vectors such that,
\begin{equation}
\mathbf{S}_n^\kk = \sum_{m,p} c_{mp} \psi^{\kk}_{\nu mp},
\end{equation}
where coefficients $c_{mp}$ can be complex. We label $\nu$ as the active irreducible representation $\Gamma_\nu$, $m=1\ldots n_\nu$, where $n_\nu$ is the number of times $\Gamma_\nu$ is contained in $\Gamma_{\rm mag}$. The index $p$ labels the component corresponding to the dimension of $\Gamma_\nu$.

\begin{figure*}
\includegraphics[bb=40 50 350 180, clip, width=0.9\textwidth]{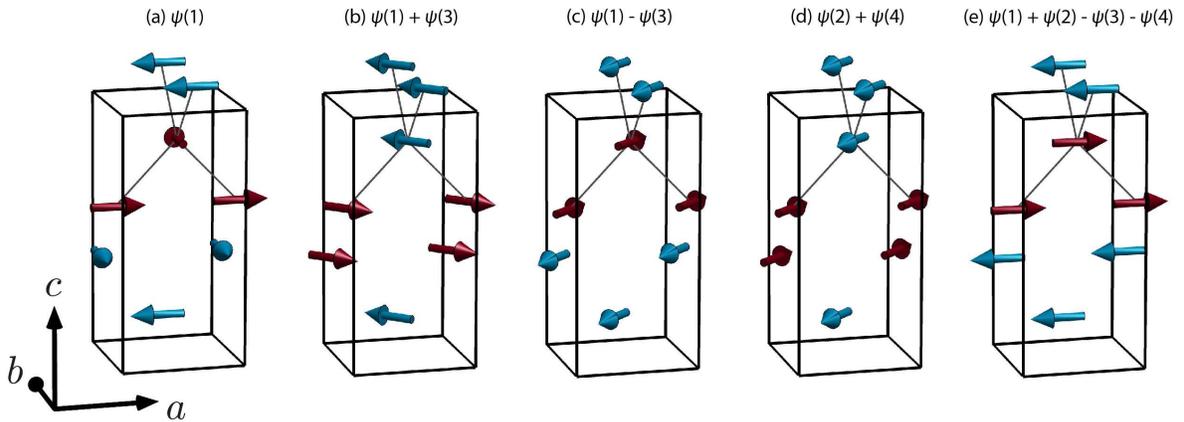}
\caption{(Color online) Possible magnetic structures of $\Gamma_1$ irreducible representation. (a) Magnetic structure from just the first basis vector in Table~\ref{tab:basis_vectors}. (b) -- (d) show arrangement of moments by combining two basis vectors. (e) Appropriate sum of all basis vectors to form magnetic structure which best describes \LEF.}
\label{figS3}
\end{figure*}

In the case of \LYF, the moments lie in the $ab$ plane, therefore $\Gamma_1$ is active (see Table~\ref{tab:basis_vectors}). However, the neutron data does not allow us to uniquely identify the magnetic ordering as any of the four basis vectors can refine the measured data. All four arrangements result in moments which rotate by 90$^\circ$ along $c$, as for example shown in Fig.~\ref{figS3}(a). It is also possible to use a combination of two basis vectors, such as 1 and 3 or 2 and 4 to describe a collinear magnetic structure as shown in Figs.~\ref{figS3}(b--d). However, it is not possible to refine the measured data for \LEF\ using the same combination of basis vectors which appear to describe \LYF. Indeed a combination of all four basis vectors, as depicted in Fig.~\ref{figS3}(e), is needed to describe the best possible solution for \LEF\ reported in Ref.~\cite{kraemer-science-2012} which sees the moments along the $a$ axis

\subsection{Crystal field interaction}

\begingroup
\begin{table}
\centering
\begin{tabular}{ l c c c c c c}
\hline
ion   & $10^3B^0_2$   & $10^3B^0_4$   & $10^6B^0_6$
         & $10^3B^4_4$(c)  & $10^3B^4_6$(c) & $10^6|B^4_6$(s)$|$\\
\hline
Er      & 58.1     & -0.536      &-0.00625
        &\textbf{-5.53}    & -0.106   &23.8        \\
        &(3.4)      & (0.032)   & (0.00041)
        & (0.31) & (0.0061)     & (1.5)          \\
Yb      &  457  & 7.75     &  0
        & \textbf{196}  & -9.78  & 0              \\
        &  (5.2)    & (0.12)     &  (0)
        & (0.65)     & (0.0094)     & (0)            \\
\hline
\end{tabular}
\caption{Crystal field parameters of \LYF\ and \LEF\ compounds determine by inelastic neutron scattering. Typically, a coordinate system with $B^4_4(s) = 0$ is chosen, while two possible equivalent coordinations of $R$ ion by F ions give different sign of $B_6^4(s)$. After \cite{babkevich-prb-2015}.
\label{tab:cf_params}}
\end{table}
\endgroup

The in-plane anisotropy in \LEF\ and \LYF\ is largely determined by the single-ion crystal field and dipolar interactions. We would expect that as the magnetic moment size is very similar in \LEF\ and \LYF, the dipolar interactions in the two systems do not differ significantly. One possible arrangement in \LYF\ is shown in Fig.~\ref{figS3}(c) where the moments are all rotated by 45$^\circ$ in the basal plane with respect to the \LEF\ magnetic structure. This structure (amongst others) fits well the measured data. From the crystal field whose Hamiltonian for $\bar{4}$ point group symmetry at the $R$ site is given by,
\begin{equation}
\mathcal{H}_{\rm CEF} = \sum_{l =2,4,6}  B_l^0{\mathbf O}_l^0 +
\sum_{l = 4,6}  B_l^4(c){\mathbf O}_l^4(c) + B_l^4(s){\mathbf O}_l^4(s).
\label{eq:crystal_field}
\end{equation}
The later $B_l^4$ terms play a role in the planar anisotropy where $B_4^4(c)$ term is found from experiments to be largest, see Table~\ref{tab:cf_params}. Classically, one obtains the energy of rotating a moment of size $J_0$ in the plane by angle $\phi$ to be $E = J_0 B_4^4(c) \cos(4\phi)$. Hence, the minimum in energy for different signs of $B_4^4(c)$ is found to be 45$^\circ$ apart. While this appears to be a simple explanation for preferred moment direction, a strong crystal field interaction would result in an Ising-like system, which is not what we observe experimentally. Furthermore, dipolar interactions are not expected to favor such ordering. Therefore, further theoretical work is necessary to examine the mechanism by which the dipolar-coupled antiferromagnets order.

\subsection{Conclusion}

While it is entirely possible that the diffraction patterns can be also described by other models including ones where moments are non-collinear, qualitatively our experimental data appears to suggest that the groundstate magnetic structure of \LEF\ is not the same as \LYF. Thus, this highlights the universality of antiferromagnetism on a distorted diamond lattice described in our Letter.

\end{document}